\documentclass[12pt]{iopart}

\begin{document}

\title[Influence of quantum energy equation on electronic plasma oscillations]{Influence of quantum energy equation on electronic plasma oscillations}

\author{A Yu Ivanov, L S Kuz'menkov}

\address{Faculty of Physics, Moscow State University, Leninskie Gory build. 1 struct. 2, Moscow, 119991, Russian Federation}
\ead{alexmax1989@mail.ru}
\begin{abstract}
Five-moment approximation in hydrodynamics includes not only continuity equation and momentum balance equation, but also energy equation. Set of hydrodynamic equations is presented for system of non-relativistic quantum particles with Coulomb interaction. This set of equations is linearized, and dispersion relation for Langmuir waves in quasi-neutral system of electrons and immobile ions is obtained.
\end{abstract}

\section{Introduction}

Investigations on quantum hydrodynamics and quantum plasma are experienced an ascension in the past few years. Problem consists in obtaining of quantum continual equations for many-particle system and investigation of known (and probably new) phenomena using this approach. Many well-known phenomena, such as propagation modes, instabilities \cite{Bret 2007}, nonlinear waves \cite{Misra Bhomwik}, ion-acoustic waves \cite{Haas Garcia et al} etc. were investigated in quantum way.

Equations of quantum plasma were studied in both kinetic and hydrodynamic approaches. Hydrodynamical approach in quantum theory was proposed by E. Madelung right after discovery of the Schr\"{o}dinger equation in 1926 \cite{Madelung}. Five-moment set of hydrodynamical equations includes evolution equation for densities of number, momentum and energy. The quantum momentum balance equation differs from classical by additional quantum pressure (this is main difference in non-relativistic case), which is not dependent from temperature and associated with Heisenberg uncertainty principle. These equations can be obtained in different ways, for example, with method described in section \ref{method}. In another way these equations can be obtained by integration of corresponding kinetic equation \cite{LSK 2002}, \cite{Haas Zamanian et al 2010}. Another prominent direction of investigations is studying of spin systems. Spin effects and contributions in equations have been discussed in \cite{LSK 2001}, \cite{Brodin Marklund 2007}, \cite{Asenjo Munoz PhysPlasm}.

Quantum effects can play significant role at high densities, when de Broglie wavelength is comparable with mean distance between particles, and also at low temperatures. Quantum equations for many-particle systems may be applied in solid-state physics \cite{Kittel}, for compact objects in astrophysics \cite{Benvenuto De Vito} and in other fields. But even at low temperatures temperature contribution may be important in the background of quantum effects. Hence, evolution equation for energy should be accounted. In current paper dispersion relation for longitudinal waves in plasma is obtained with presence of energy equation. Energy equation represents, in essence, the main equation of non-equilibrium thermodynamics; it corresponds to the equilibrium theory in absence of time dependence and currents in a system. Electronic oscillations in plasma are important theme of investigations from long ago. In fact, they are oscillations only in case of longitudinal excitations in classical cold plasma, when frequency does not depend from wave vector and, consequently, group velocity of this excitations is equal to zero. Investigation of Langmuir waves in plasma remains the actual theme up to now; for example, in \cite{Ivanov Andreev 2013} and \cite{Asenjo et al NJP 2012} waves in semi-relativistic plasma described by Breit Hamiltonian are discussed. Other wave phenomena are studied, for example, waves generated by neutron beam in plasma \cite{Andreev IntJMP 2012}, waves in polarized two-dimensional systems \cite{Andreev Kuzmenkov PRB}.

\section{The method}\label{method}

Let's consider quantum system of N particles with Coulomb interaction; this system is placed into external electromagnetic field. Derivation of quantum hydrodynamical equations is carried out by the method presented in \cite{LSK 1999}. Microscopic number density is defined as
\begin{equation}\label{definition_of_n}n(\mathbf{r},t)=\int{dR}\sum_{i=1}^N \delta(\mathbf{r}-\mathbf{r}_i)\psi^*(R,t)\psi(R,t),\end{equation}
where $\psi(R,t)$ is a wave function of the system, $R=(\mathbf{r}_1,...,\mathbf{r}_N)$, $\mathbf{r}_i$ is the radius vector of particle,
\begin{equation}dR=\prod_{j=1}^N d\mathbf{r}_j,\end{equation}
$dR$ is element of volume in 3N configured space, $d\mathbf{r}_j$ is element of volume in 3D space of vector $\mathbf{r}_j$.

In quantum mechanics state of system is described by wave function defined in 3N configured space. Quantum hydrodynamical description consists in transition from the wave function in configured space to the functions in physical 3D space, such as number density $n(\mathbf{r},t)$, velocity field $v_\alpha(\mathbf{r},t)$, kinetic pressure $p_{\alpha\beta}(\mathbf{r},t)$ etc. These functions are moments of distribution function in kinetic approach, so approximation including equations for number density, velocity field and energy density is called five-moment approximation.

Differentiating (\ref{definition_of_n}) over $t$ and using Schr\"{o}dinger equation for N particles with Coulomb interaction, it is obtained first equation in a set of equations, and it is continuity equation. In this equation new quantity appears, which represents current density. For current density it can be derived next equation, which is momentum balance equation. Full set of hydrodynamical equation can be obtained in this way.

\section{Set of equations}\label{set_of_eq}

Let's consider a system with one kind of particles (i. e. $m_i=m$, $e_i=e$ for each $i$). Using method of derivation described in section \ref{method}, set of hydrodynamical equations for quantum system of particles with the Coulomb interaction can be obtained \cite{LSK 1999}. The continuity equation have next form:
\begin{equation}\label{continuity_eq}\partial_t n+\mathrm{div}(n\mathbf{v})=0,\end{equation}
where $n\mathbf{v}=\mathbf{j}$ is number current density, $\mathbf{v}$ is velocity field. The momentum balance equation is
\begin{equation}\label{momentum_bal_eq}mn(\partial_t+v_\beta\partial_\beta)v_\alpha+
\partial_\beta p_{\alpha\beta}+\partial_\beta T_{\alpha\beta}=enE_\alpha+\frac{e}{c}\varepsilon_{\alpha\beta\gamma}nv_\beta B^{ext}_\gamma,\end{equation}
where $p_{\alpha\beta}$ is kinetic pressure tensor, $T_{\alpha\beta}$ is additional quantum pressure, which corresponds to the quantum Bohm potential. These tensor have next form:
\begin{equation}T_{\alpha\beta}(\mathbf{r},t)=-\int dR\sum_{i=1}^N\delta(\mathbf{r}-\mathbf{r}_i)\frac{\hbar^2}{2m}a^2
\partial_{i\alpha}\partial_{i\beta}\ln a,\end{equation}
where $a$ is an amplitude of wave function $\psi(R,t)=a\exp(iS(R,t)/\hbar)$. In the approximation of weakly interacting particles ($\psi(R,t)=\psi(\mathbf{r}_1)\psi(\mathbf{r}_2)...\psi(\mathbf{r}_N)$) the quantum pressure $\partial_\beta T_{\alpha\beta}$ leads to the term in the form of
\begin{equation}\partial_\beta T_{\alpha\beta}=-\frac{\hbar^2}{2m}\partial_\alpha\biggl(\frac{\triangle\sqrt{n}}{\sqrt{n}}
\biggr{)}\end{equation}
in the momentum balance equation (\ref{momentum_bal_eq}). It needs to be done to close the set of equations.
The energy balance equation is
\begin{eqnarray}\label{energy_eq}
&&n(\partial_t+v_\beta\partial_\beta)\epsilon+(p_{\alpha\beta}+T_{\alpha\beta})\partial_\alpha v_\beta+\partial_\alpha q_\alpha=\nonumber\\
&&{}=-\frac{e^2}{2}n\int d\mathbf{r}'(v_\alpha(\mathbf{r}',t)-v_\alpha(\mathbf{r},t))\partial_\alpha G(\mathbf{r}-\mathbf{r}')n(\mathbf{r}',t)+\alpha,
\end{eqnarray}
where $n\epsilon$ is thermal energy density, $q_\alpha$ is vector of thermal energy flux, $\alpha$ is work density, $G(\mathbf{r}-\mathbf{r}')=1/|\mathbf{r}-\mathbf{r}'|$ is the Green function of Coulomb interaction. Functions $n\epsilon$ and $q_\alpha$ have next form:
\begin{eqnarray}
&n\epsilon&=\int{dR}\sum_{i=1}^N \delta(\mathbf{r}-\mathbf{r}_i)a^2\biggl(\frac{1}{2}m u_i^2-\frac{\hbar^2}{2m}\frac{\triangle_i a}{a}\biggr)+\nonumber\\
&&{}+\frac{e^2}{2}n\int{d\mathbf{r}'}G(\mathbf{r}-\mathbf{r}')n(\mathbf{r}',t),
\end{eqnarray}
\begin{eqnarray}\label{thermal_energy_flux}
&q_\alpha&=\int{dR}\sum_{i=1}^N \delta(\mathbf{r}-\mathbf{r}_i)a^2\biggl[u_{i\alpha}\biggl(\frac{1}{2}m u_i^2-\frac{\hbar^2}{2m}\frac{\triangle_i a}{a}\biggr)-\nonumber\\
&&{}-\frac{\hbar^2}{2m}\partial_{i\alpha}(v_{i\beta}\partial_{i\beta}\ln{a})-
\frac{\hbar^2}{4m}\partial_{i\alpha}\partial_{i\beta}v_{i\beta}\biggr]+\nonumber\\
&&{}+\frac{1}{2}\int{d\mathbf{r}'}e^2G(\mathbf{r}-\mathbf{r}')\int{dR}\sum_{i,j=1,j\neq i}^N \delta(\mathbf{r}-\mathbf{r}_i)\delta(\mathbf{r}'-\mathbf{r}_j)a^2u_{i\alpha},
\end{eqnarray}
where $u_{i\alpha}$ is thermal velocity of particle, $u_{i\alpha}=v_{i\alpha}-v_\alpha$. Equations (\ref{continuity_eq}), (\ref{momentum_bal_eq}) and (\ref{energy_eq}) form five-moment set of hydrodynamical equations. This set must be closed for applying to specific problems. First, the equation of state $p=(2/3)n\epsilon$ should be used (kinetic pressure tensor is diagonal: $p_{\alpha\beta}=p\delta_{\alpha\beta}$), which is direct consequence of pressure tensor and energy density definitions for isotropic case. Second, we should neglect the thermal fluxes of Coulomb interaction energy and work density. Third, Fourier law should be applied for classical part of thermal energy flux:
\begin{equation}
\int{dR}\sum_{i=1}^N \delta(\mathbf{r}-\mathbf{r}_i)a^2u_{i\alpha}\frac{1}{2}m u_i^2=-\frac{3}{2}n\kappa\partial_\alpha T.
\end{equation}
In other words, it is possible to transfer $u_{i\alpha}$ over the integration limits and to substitute it on operator $-\kappa\partial_\alpha$ ($\kappa$ is coefficient, which is thermal conductivity, number coefficient $3/2$ was added for correspondence with well-known heat equation; $T$ is local value of temperature). Quantum terms in thermal energy flux (\ref{thermal_energy_flux}), proportional to $\hbar$, can be closed in the same way.

\section{Linearized equations and dispersion relation for Langmuir waves}

Let's consider quasi-neutral system of electrons and immobile ions. Linearized set of equations for this case have next form:
\begin{eqnarray}
\partial_t n'+n_0\mathrm{div}\mathbf{v}'=0,
\end{eqnarray}
\begin{eqnarray}&&mn_0\partial_t v'_\alpha-\frac{\hbar^2}{4m}\partial_\alpha\triangle n'+T_0\partial_\alpha n'+n_0\partial_\alpha T'=\nonumber\\
&&{}=-e^2n_0\partial_\alpha\int{d\mathbf{r}'}G(\mathbf{r}-\mathbf{r}')n'(\mathbf{r}',t),
\end{eqnarray}
\begin{eqnarray}
&&n_0\partial_t\biggl(\frac{3}{2}T'+\frac{e^2}{2}
\int{d\mathbf{r}'}G(\mathbf{r}-\mathbf{r}')n'(\mathbf{r}',t)\biggr)+n_0 T_0\mathrm{div}\mathbf{v}'-\nonumber\\
&&{}-\frac{3}{2}\kappa n_0\triangle T'+\frac{\hbar^2}{4m}\kappa_q\triangle\triangle n'=-\frac{e^2}{2}n_0^2\int{d\mathbf{r}'}\partial_\alpha G(\mathbf{r}-\mathbf{r}')v'_\alpha(\mathbf{r}',t),
\end{eqnarray}
where $n=n_0+n'$ is number density of electrons ($n_0$ is equilibrium density, $n'$ is perturbative density; analogous designations are applied for other quantities). Proportional to constant coefficient $\kappa_q$ term appears at closing quantum part of thermal energy flux (in the way decribed in the end of section \ref{set_of_eq}).

Carrying Fourier transformation, a set of linear algebraic equations is obtained. It is possible to get next dispersion equation for waves in this system:
\begin{equation}\label{disp_eq}
(\omega+ik^2\kappa)\biggl(\omega^2-\omega^2_p-\frac{\hbar^2k^4}{4m^2}-\frac{T_0k^2}{m}\biggr)-
\frac{2}{3}\frac{T_0k^2}{m}\omega+i\kappa_q\frac{\hbar^2k^6}{6m^2}=0,
\end{equation}
where $\omega_p^2=4\pi e^2n_0/m$ is Langmuir frequency. This is third order algebraic equation respectively $\omega(k)$. General solution for this equation is given by Cardano formulas. This formulas give solution in rather uncertain form, so let's apply it for the most practical case when $\omega^2_p+\frac{T_0k^2}{m}$ is bigger than any other coefficients in (\ref{disp_eq}). Dispersion branches have next form:
\begin{equation}\label{omega_1}
\omega_1(k)=-\frac{1}{3}i\kappa k^2+\frac{\sqrt{3}}{2}(A+B)+i\frac{1}{2}(A-B),
\end{equation}
\begin{equation}\label{omega_2,3}
\omega_{2,3}(k)=-\frac{1}{3}i\kappa k^2+\frac{\sqrt{3}}{4}(-1\mp 1)(A+B)+i\frac{1}{4}(-1\pm 3)(A-B),
\end{equation}
where $A(k)=(\frac{q}{2}+\sqrt{\frac{p^3}{27}+\frac{q^2}{4}})^{\frac{1}{3}}$, $B(k)=(-\frac{q}{2}+\sqrt{\frac{p^3}{27}+\frac{q^2}{4}})^{\frac{1}{3}}$. Functions $p=p(k)$ and $q=q(k)$ define in next way:
\begin{equation}\label{p}
p=\omega_p^2+\frac{\hbar^2k^4}{4m^2}+\frac{5T_0k^2}{3m}-\frac{1}{3}(\kappa k^2)^2,
\end{equation}
\begin{equation}\label{q}
q=\kappa k^2\biggl[\frac{2}{27}(\kappa k^2)^2+\frac{2}{3}\omega_p^2+\frac{4T_0k^2}{9m}+\frac{\hbar^2k^4}{6m^2}\biggr]-\kappa_q
\frac{\hbar^2k^6}{6m^2}.
\end{equation}

In ordinary consideration, when energy equation is not presented, dispersion equation is second order algebraic equation. It has very simple form: $\omega^2-\omega^2_p-\frac{T_0k^2}{m}-\frac{\hbar^2k^4}{4m^2}=0$. There are two branches, which differ only by the sign. The situation is changed in the case described above. Branches $\omega_{1,2}(k)$ corresponding to usual branches have imaginary part and differ not only by the sign. Also, there is new branch $\omega_{3}(k)$, which is totally imaginary in the presented case. Imaginary parts correspond to the damping of waves.

Formulas (\ref{omega_1}) and (\ref{omega_2,3}) have quite awkward form, so let's consider case when $\omega_p^2$ much more than other terms in (\ref{p}), (\ref{q}). Let's expand radical expressions in (\ref{omega_1}), (\ref{omega_2,3}) using Taylor series to the second order of $\omega_p^{-1}$. Using this approach more simple expressions for dispersion branches can be obtained:
\begin{eqnarray}\label{omega_1,2}
&&\omega_{1,2}(k)=\pm\omega_p\biggl[1+\frac{1}{2\omega_p^2}\biggl(\frac{\hbar^2k^4}{4m^2}+
\frac{5T_0k^2}{3m}+\frac{2}{3}(\kappa k^2)^2\biggr)\biggr]+\nonumber\\
&&{}+\frac{i}{2}\kappa k^2\biggl[\frac{2}{3}+\frac{1}{\omega_p^2}\biggl(\frac{2}{27}(\kappa k^2)^2+\frac{4T_0k^2}{9m}+\frac{\hbar^2k^4}{6m^2}-\frac{\kappa_q}{\kappa}
\frac{\hbar^2k^4}{6m^2}\biggr)\biggr],
\end{eqnarray}
\begin{equation}
\omega_{3}(k)=-i\kappa k^2\biggl[\frac{2}{3}+\frac{1}{\omega_p^2}\biggl(\frac{2}{27}(\kappa k^2)^2+\frac{4T_0k^2}{9m}+\frac{\hbar^2k^4}{6m^2}-\frac{\kappa_q}{\kappa}
\frac{\hbar^2k^4}{6m^2}\biggr)\biggr].
\end{equation}

So, looking on this expressions it is easy to conclude that branches $\omega_{1,2}$ indeed correspond to well-known branches, and $\omega_{3}$ is essentially new branch. Note that number coefficient before temperature term in (\ref{omega_1,2}) equals to $5/3$, which differs from $1$ in isothermal case and $3$ in one-dimensional adiabatic case \cite{Krall Trivelpiece}. In this paper consideration is similar to isothermal case, but additional number coefficient $2/3$ arises form term $p\partial_\alpha v_\alpha$ in the energy equation (\ref{energy_eq}).

\section{Conclusion}

In the beginning of this paper brief description of quantum hydrodynamic method is presented. Equations of many-particle quantum hydrodynamics can be obtained for Hamiltonian systems; in this paper five-moment set of hydrodynamic equations is presented for quantum system of particles with Coulomb interaction. External electromagnetic field also can be taken into account. This set of equations is linearized and dispersion relation for Langmuir waves in plasma is obtained. Only self-consistent field approximation is considered. The energy equation, unlike the momentum equation, does not contain quantum non-thermal terms in linear case. Quantum thermal term in energy equation are closed using analogy with Fourier law for classical thermal flux. Unlike the usual case, there are three dispersion branches, one of which is new. New branch is totally imaginary. There is energy damping due to dissipation of thermal energy and tendency to the equilibrium state.

\


\begin{thebibliography}{99}

\bibitem{Bret 2007} A. Bret. Phys. Plasm. 2007. V. 14. P. 084503.

\bibitem{Misra Bhomwik} A. P. Misra, C. Bhowmik. 2007. Phys. Plasm. V. 14. P. 012309.

\bibitem{Haas Garcia et al} F. Haas, L. G. Garcia, J. Goedert, G. Manfredi. Phys. Plasm. 2003. V. 10. P. 3858.

\bibitem{Madelung} E. Madelung. Z. Phys. 1926. V. 40. P. 332.

\bibitem{LSK 2002} L. S. Kuzmenkov, S. G. Maksimov. Theor. Math. Phys. 2002. V. 131(2). P. 641.

\bibitem{Haas Zamanian et al 2010} F. Haas, J. Zamanian, M. Marklund, G. Brodin. New J. Phys. 2010. V. 12. P. 073027.

\bibitem{LSK 2001} L. S. Kuzmenkov, S. G. Maksimov, V. V. Fedoseev. Theor. Math. Phys. 2001. V. 126(1). P. 110.

\bibitem{Brodin Marklund 2007} G. Brodin, M. Marklund. New J. Phys. 2007. V. 9. P. 277.

\bibitem{Asenjo Munoz PhysPlasm} F. A. Asenjo, V. Mu\~{n}oz, J. A. Valdivia, S. M. Mahajan. Phys. Plasm. 2011. V. 18. P. 012107.

\bibitem{Kittel} C. Kittel. Introduction to Solid State Physics. John Wiley \&
Sons Inc., New York, 1996.

\bibitem{Benvenuto De Vito} O. G. Benvenuto, M. A. De Vito. Mon. Not. R. Astron.
Soc. 2005. V. 362. P. 891.

\bibitem{Ivanov Andreev 2013} A. Yu. Ivanov, P. A. Andreev. Rus. J. Phys. 2013. V. 56. N. 3. P. 325.

\bibitem{Asenjo et al NJP 2012} F. A. Asenjo, J. Zamanian, M. Marklund, G. Brodin, P. Johansson. New J. Phys. 2012. V. 14. P. 073042.

\bibitem{Andreev IntJMP 2012} P. A. Andreev, L. S. Kuz'menkov. Int. J. Mod. Phys. B. 2012. V. 26. N. 32. P. 1250186.

\bibitem{Andreev Kuzmenkov PRB} P. A. Andreev, L. S. Kuz'menkov, M. I. Trukhanova. Phys. Rev. B. V. 84. P. 245401.

\bibitem{LSK 1999} L. S. Kuzmenkov, S. G. Maksimov. Theor. Math. Phys. 1999. V. 118(2). P. 227.

\bibitem{Krall Trivelpiece} N. A. Krall, A. W. Trivelpiece. Principles of Plasma Physics. San Francisco Press, 1986.

\end{thebibliography}
\end{document}